\documentclass[aps,prl,floatfix,twocolumn,amsmath,amssymb,nofootinbib]{revtex4-1}
\usepackage{amsthm,color,latexsym,graphicx,array,multirow,epstopdf}
\usepackage[hyperfootnotes=true]{hyperref}

\definecolor{darkred}{rgb}{0.5,0.0,0.0}
\definecolor{darkblue}{rgb}{0.0,0.0,0.9}
\definecolor{darkerblue}{rgb}{0.0,0.0,0.5}
\definecolor{darkgreen}{rgb}{0.0,0.5,0.0}
\definecolor{black}{rgb}{0.0,0.0,0.0}
\definecolor{brown}{rgb}{0.6,0.4,0.2}

\def\be{\begin{equation}}
\def\ee{\end{equation}}

\def\cO{\mathcal{O}}

\def\msbar{\overline{\text{MS}}}

\def\muX{\mu_X}

\def\nn{\nonumber}

\newcommand{\vew}{v_{\text{EW}}}
\newcommand{\Vew}{V_{\text{EW}}}
\newcommand{\Vmin}{V_{\text{min}}}
\newcommand{\Vmax}{V_{\text{max}}}
\newcommand{\VLO}{V^{\text{(LO)}}}

\newcommand{\GeV}{\text{GeV}}
\newcommand{\mhpole}{ {m_h^{\text{pole}}}}
\newcommand{\mtpole}{ {m_t^{\text{pole}}}}

\newcommand{\mpl}{  M_{\text{Pl}} }
\newcommand{\LNP}{ \Lambda_{\text{NP}} }

\begin{document}

\title{Consistent Use of the Standard Model Effective Potential}
\author{Anders Andreassen}
\email{anders@physics.harvard.edu} 
\affiliation{Department of Physics, Harvard University, Cambridge, MA 02138, USA}
\author{William Frost}
\email{wfrost@physics.harvard.edu} 
\affiliation{Department of Physics, Harvard University, Cambridge, MA 02138, USA}
\author{Matthew D. Schwartz}
\email{schwartz@physics.harvard.edu}
\affiliation{Department of Physics, Harvard University, Cambridge, MA 02138, USA}

\begin{abstract}
The stability of the Standard Model is determined by the true minimum of the effective Higgs potential. 
We show that the potential at its minimum when computed by the traditional method is strongly
dependent on the gauge parameter. It moreover depends on the scale where the potential is calculated. We provide a consistent method for determining absolute stability independent of both gauge 
and calculation scale, order by order in perturbation theory. This leads to a revised stability bounds
$\mhpole > (129.4 \pm 2.3)~\GeV$
and $\mtpole < (171.2 \pm 0.3)~\GeV$.
We also show how to evaluate the effect of new physics on the stability bound without resorting to unphysical field values. 
\end{abstract}

\maketitle

An intriguing  consequence  of the  recent discovery of the Higgs boson is that its mass apparently places the Standard Model (SM) near the border between absolute stability and metastability~\cite{Degrassi:2012ry,Buttazzo:2013uya}. A renowned plot,  
Fig.~3 of~\cite{Buttazzo:2013uya},
 shows the standard model lying close to the end of a metastability funnel in the Higgs-mass/top-mass plane. This unanticipated tuning has inspired a fair amount of speculation about its possible origin and implications.  
Stability is normally determined by examining the zero-temperature effective potential $V$ for the SM~\cite{Cabibbo:1979ay,Sher:1988mj,Casas:1994qy,Degrassi:2012ry,Buttazzo:2013uya}: if this potential has a negative minimum at large field values, the SM is said to be unstable; if the inverse decay rate for tunneling out of the electroweak minimum is larger than the lifetime of the universe, the SM is said to be metastable.
While these criteria are physical, the  extraction of numerical bounds within a consistent perturbation expansion is not straightforward. 

One complication in making physical predictions with $V$ is that effective potentials are not gauge-invariant~\cite{Jackiw:1974cv}. 
Although physical quantities extracted from an effective potential (or more generally from an effective action) must be gauge-invariant, 
 there have been surprisingly few explicit 
 checks~\cite{Kang:1974yj,Dolan:1974gu,Aitchison:1983ns,Loinaz:1997td}.
The traditional approach is simply to work in Landau gauge where calculations are easiest and to {\it assume} that the approximations used are self-consistent.

 Progress in understanding the gauge-dependence  was made by Nielsen~\cite{Nielsen:1975fs} and independently by Kugo and Fukuda~\cite{Fukuda:1975di} in 1975. One result from these papers is that the effective potential satisfies a differential equation:
\be
\left( \xi \frac{\partial}{\partial \xi} + C(h,\xi) \frac{\partial}{\partial h} \right) V(h,\xi) =0 \label{Nielsen}
\ee
where $\xi$ is the gauge parameter in Fermi gauges and $C(h,\xi)$ is a calculable function. This {\it Nielsen identity} says that the gauge-dependence of the effective potential can be compensated for by a rescaling of the scalar field $h$. Two generic implications are that 1) the value of the field $h$ can never be physical, since any rescaling of the field can be compensated by a gauge-change and 2) the value of $V(h,\xi)$ at an extremum in $h$ should be gauge-invariant. Gauge-dependent quantities then include the value of $V$ at any non-extremal point, and the value of $h$ at {\it any} point (extremal or not). It is worth noting that Eq.~\eqref{Nielsen} is not quite as powerful as it might seem, since $C(h,\xi)$ can be infinite in perturbation theory~\cite{Nielsen:1975fs,Aitchison:1983ns,Andreassen:2014eha}.

Despite the widespread contentment with Landau gauge, the gauge dependence of the effective potential has occasionally caused some discomfort~\cite{DiLuzio:2014bua,Boyanovsky:1997dj,Wainwright:2011qy,Cook:2014dga,Andreassen:2013hpa}. A handful of papers have proposed field redefinitions to generate a gauge-independent potential~\cite{Frere:1974ia,Tye:1996au,Nielsen:2014spa}. 
 This approach purportedly allows the effective potential to be used like a classical potential, assigning physical significance to both  field values and the potential at each point. 
 However, it is not clear why removing the gauge-dependence automatically makes the potential physical. Moreover, for the field redefinition to be justified it should leave physical quantities unchanged; in that case, we may  as well work with fields that make the calculations easiest.

The effective potential has another feature which has not generally been appreciated: it depends on the scale where it is calculated. To see this, note that $V$ satisfies a renormalization group equation~\cite{Coleman:1973jx}:
\be
\Big(\mu \frac{\partial}{\partial \mu}
 - \gamma h \frac{\partial}{\partial h} 
 + \beta_i \frac{\partial}{\partial \lambda_i}\Big) V= 0
\label{RGE}
\ee
This equation says that the explicit $\mu$-dependence of the potential can be compensated for by rescaling the couplings according to their $\beta$-functions and rescaling the field $h$ according to its anomalous dimension, $\gamma$. Thus, if we know the potential at a scale $\mu_0$ in terms of the couplings $\lambda_i (\mu_0)$ we can find it at a scale $\mu$ by solving this equation. Call this method 1. Alternatively, we could have just computed it at the scale $\mu$ to begin with, in terms of the couplings $\lambda_i(\mu)$. Call this method 2. Methods 1 and 2 do not give the same potential, to any order or to all orders in perturbation theory. They differ by the rescaling of the field $h$.  This is not a problem, since we have already concluded from Eq.~\eqref{Nielsen} that physical quantities extracted from $V$ should be independent of field rescaling. 
The additional freedom of choosing $\mu_0$ illustrates that even in a fixed gauge or with gauge-invariant composite fields, field values are still unphysical. 

The unphysical nature of $V$ may be less unsettling after
 recalling that the effective potential is the constant-field limit of the 1PI effective action. The vertices of this action at tree-level produce 1PI correlation functions, which can be gauge and scale dependent and satisfy an RGE like Eq.~\eqref{RGE}. Gauge-invariant $S$-matrix elements are related to correlation functions by amputation and (in $\msbar$) gauge-dependent wave-function renormalization $Z$-factors. These factors also compensate the scale dependence, letting the $S$-matrix satisfy an RGE like Eq.~\eqref{RGE} without the $\gamma$ term. 

\begin{figure}[t]
   \includegraphics[width=0.43\textwidth]{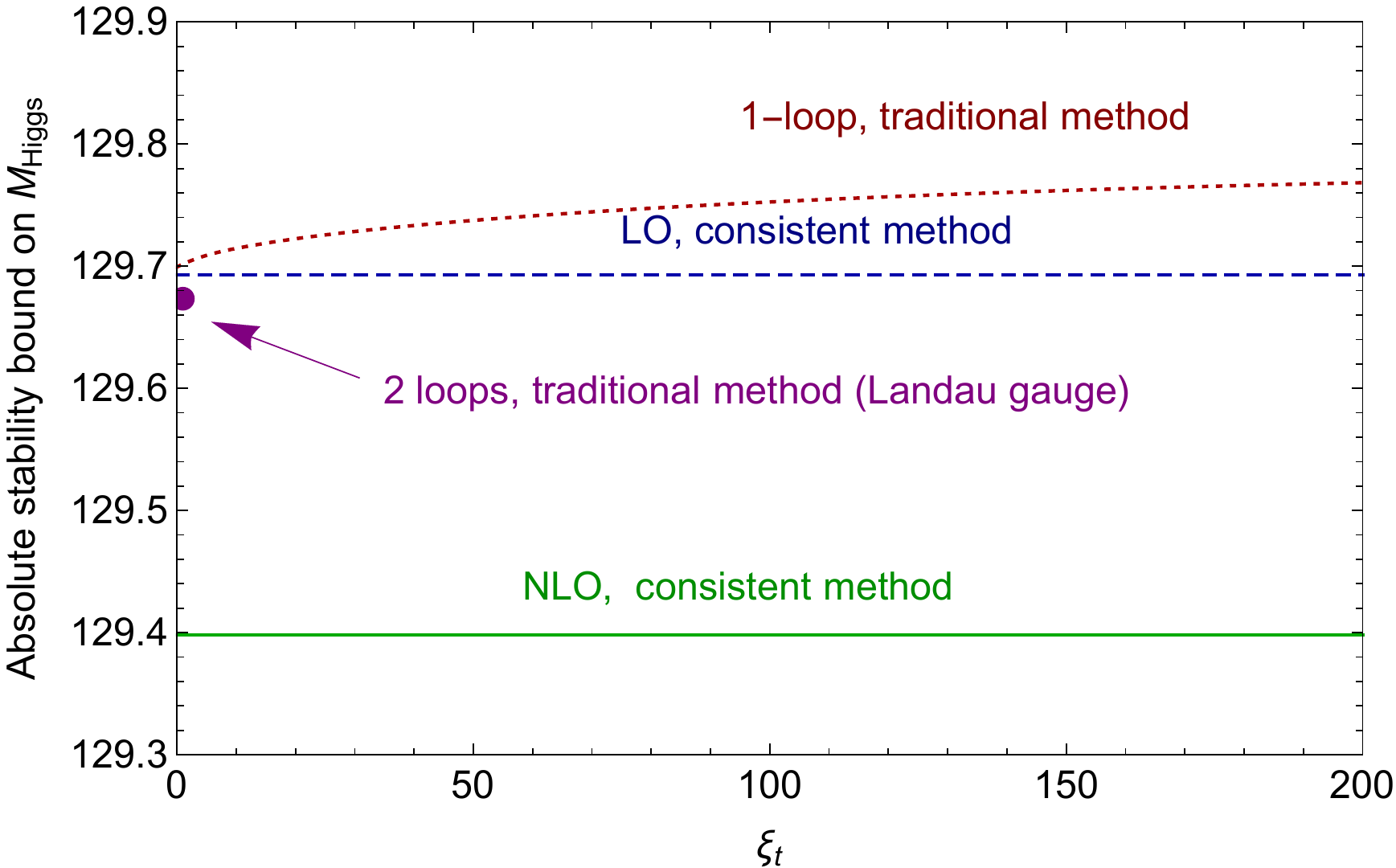}
\caption{Gauge dependence of the absolute stability bound with $\mtpole = 173.34~\GeV$.}
 \vspace{-3ex}
  \label{fig:massbound}
 \end{figure}

Fortuitously,  the value of the  potential at a minimum, $\Vmin$, (or at any extremum) is both gauge-invariant and independent of the scale where it is calculated, {\it without} extra $Z$-factors. The former invariance follows from Eq.~\eqref{Nielsen}
 and the latter invariance holds
simply because the value of any function at any extremum is invariant under  any rescaling of its argument. Since the absolute stability bound in the SM is determined by the condition $\Vmin<\Vew\approx 0$, with $\Vew$ the  energy of our vacuum (usually renormalized to zero), the bound should be gauge-independent. Unfortunately, gauge-invariance has only been proven non-perturbatively. Indeed, we find that the stability bound is  gauge-dependent at each order in perturbation theory if computed by the traditional approach (see Fig.~\ref{fig:massbound} or~\cite{Andreassen:2013hpa}).  In~\cite{Andreassen:2014eha}, it was shown how effective potential calculations can be reorganized so that $\Vmin$
is gauge-invariant order-by-order.  In this 
paper, we review this ``consistent approach'' and apply it to the SM.

We write the SM effective potential as $V(h)$, where in unitary gauge
the Higgs doublet is  normalized as
$H=\dfrac{1}{\sqrt{2}} \begin{pmatrix} 0 \\  \vew+h \end{pmatrix}$.
The traditional perturbation approach leads to a renormalization-group-improved effective potential of the form~\cite{Buttazzo:2013uya}
\be
V(h) =\frac{1}{4}h^4 e^{4\Gamma(h)} \left[ \lambda_{\text{eff}}^{(0)}(\mu = h) +  \lambda_{\text{eff}}^{(1)}(\mu = h) +   
 \cdots \right]  \label{Voldis}
\ee
with $\Gamma(h) \equiv \int_{m_t}^h \gamma(\mu') \frac{d \mu'}{\mu'}$ and  $\frac{1}{4}\lambda_{\text{eff}}^{(j)}(\mu)h^4$ the 
$j$-loop fixed-order effective potential.

 Since stability is determined by large field values and the potential grows as $h^4$, the quadratic term $-m^2 h^2$ in the classical potential  can be neglected to excellent accuracy. Then the electroweak minimum is at $\Vew = 0$ and the stability bound is determined as the critical Higgs pole mass for which the potential has another minimum with $\Vmin=0$. The physical  Higgs mass enters through threshold corrections at the weak scale which convert observables into $\msbar$ couplings. Currently, the  $\beta$-functions and $\gamma$ are known to 3-loop order in general $R_\xi$ gauges, the fixed-order potential is known to 2-loop order in Landau gauge ($\xi=0$), and the threshold corrections are known to 2 loops
(an alternate scheme is discussed in~\cite{Spencer-Smith:2014woa}).  Using Eq.~\eqref{Voldis} and the best available data, Ref.~\cite{Buttazzo:2013uya} found an absolute stability bound of 
 $\mhpole > (129.1 \pm 1.5)~\GeV$. Using equations from~\cite{Buttazzo:2013uya}, with some minor corrections confirmed by its authors, and including tau and bottom contributions, we have reproduced this result. We now update
 the top mass to $\mtpole = (173.34 \pm 1.12)~\GeV$, with the central value and $\pm 0.76~\GeV$ of the uncertainty from~\cite{ATLAS:2014wva},
and an additional $0.82 \GeV$ theory uncertainty added in quadrature due to the ambiguity in converting from a Monte Carlo mass scheme to a pole mass scheme~\cite{Hoang:2008xm,Moch:2014tta}.
Also including the 3-loop QCD threshold corrections to $\lambda$ listed but not used in~\cite{Buttazzo:2013uya}, we update this traditional-approach bound to $\mhpole > (129.67 \pm 1.5)~\GeV$.

The gauge-dependence of the stability bound at 1-loop is shown in Fig.~\ref{fig:massbound}, to be discussed more below.
The reason the stability bound appears gauge dependent is due to an improper use of perturbation theory. The key insight, made long ago by Coleman and Weinberg~\cite{Coleman:1973jx} is that the usual loop expansion is inappropriate for effective potentials near quantum-generated minima. Simply put, 
the classical potential $V_0 \sim \lambda h^4$ can only turn over due to 1-loop corrections of the form  $V_1 \sim \frac{g^4 \hbar}{16\pi^2}h^4$ for
some  $g$
if $\lambda \sim  \frac{g^4 \hbar}{16\pi^2}$. Since $\lambda\sim \hbar$, each factor of $\lambda$ in a diagram changes its effective loop order. Thus perturbation theory in $\hbar$ may still be appropriate, but since $\lambda\sim \hbar$ it is not the usual loop expansion. 

An additional complication is that the effective potential has terms scaling like inverse powers of $\hbar$. For example, a term $\sim \hbar^3 g^{10}\lambda^{-1}$ appears at 3-loops; since $\lambda$ counts as $\hbar$, this term scales like $\hbar^2$ and contributes competitively with the 2-loop terms. Including all relevant terms according to this modified power counting, it was shown in~\cite{Andreassen:2014eha} that $\Vmin$ is indeed gauge-invariant in scalar QED. The required terms include the 2-loop effective potential in $R_\xi$ gauge as well as an infinite series of ``daisy'' loops producing terms in $V$ proportional to $g^{4j+2} \lambda^{1-j}$.  

The consistent method for an order-by-order gauge-independent calculation of $\Vmin$ presented in~\cite{Andreassen:2014eha} translates to the SM as follows.
First, we truncate the effective potential to order $\hbar$ with $\lambda \sim \hbar$ power counting. This gives the leading-order (LO) potential:
\begin{multline}
\VLO(h) = \frac{1}{4} \lambda h^4 \\
+ h^4 \frac{1}{2048 \pi^2}
\Big[-5 g_1^4+6 (g_1^2+g_2^2)^2 \ln \frac{h ^2 (g_1^2+g_2^2)}{4 \mu ^2}\\
-10 g_1^2 g_2^2-15 g_2^4+12
   g_2^4 \ln \frac{g_2^2 h ^2}{4 \mu ^2}+144 y_t^4-96 y_t^4 \ln \frac{y_t^2 h ^2}{2 \mu ^2}
\Big]
\end{multline}
Note that this potential includes tree-level and 1-loop contributions, and is gauge-invariant.
From this, we can solve for the  scale $h=\muX$ where $d \VLO/dh = 0$.  Explicitly
$\muX$ is the $\msbar$  scale where the condition
\begin{multline}
\lambda = \frac{1}{256 \pi^2}\Big[
g_1^4+2 g_1^2 g_2^2+3 g_2^4 -48 y_t^4\\
-3 \left(g_1^2+g_2^2\right)^2 \ln \frac{g_1^2+g_2^2}{4}-6 g_2^4 \ln
   \frac{g_2^2}{4}+48 y_t^4 \ln \frac{y_t^2}{2}
   \Big] \label{lambdacond}
\end{multline}
is satisfied.
For values of $m_h$ and $m_t$ close to the observed SM values, there are two solutions to this equation: the lower $\muX$ is where $\VLO$ has a maximum, and the higher $\muX$ where the minimum occurs.  In the SM these scales are
\begin{align}
\muX^{\text{max}} &= 2.46 \times 10^{10} ~\GeV\\
\muX^{\text{min}} &= 3.43 \times 10^{30} ~\GeV.
\end{align}
These numbers and results which follow use $\mhpole = (125.14 \pm 0.24 )~\GeV$, combined from~\cite{CMSHiggs,Aad:2014aba}.

For the potential at the next-to-leading order (NLO), one contribution comes from
the $\hbar^2$ terms in the 1-loop potential with $\lambda \sim \hbar$ scaling:
\begin{multline}
V^{(1,\text{NLO}) }(h)= \frac{-1}{256 \pi^2}
 \left[ \xi_B g_1^2 \left(\ln\frac{\lambda h^4 (\xi_B g_1^2 + \xi_W g_2^2 )}{4\mu^4}-3\right) \right.\\
\left.+ \xi_W g_2^2 \left( \ln\frac{\lambda^3 h^{12}\xi_W^2 g_2^4 (\xi_B g_1^2 + \xi_W g_2^2 )}{64 \mu^{12}}- 9\right)  \right]\lambda h^4
\end{multline}
Another contribution $V^{(2,\text{NLO}) }(h)$ comes from the $\lambda^0$ and $\ln\lambda$ terms in 2-loop potential. 
In Landau gauge, these terms are $h^4/4$ times what is written as $\lambda_{\text{eff}}^{(2)}$ in Eq. (C.4) of
the published version of~\cite{Buttazzo:2013uya}. 
Finally, there is the contribution, $V^{(n>2,\text{NLO})}(h)$ from 3-loop and higher order graphs proportional to inverse powers of $\lambda$. Including all these terms, the potential at each extremum will be gauge-invariant. Conveniently, the
higher-loop-order graphs contributing at NLO vanish in Landau gauge ($\xi_B=\xi_W=0$). Thus the gauge-invariant NLO value of
the potential at the minimum is simply
\be
V_{\text{min}}^{\text{NLO}} = \VLO(\muX) + V^{(2,\text{NLO}) }(\muX) 
\label{Vminus}
\ee
To derive this, we consistently truncated to $\cO(\hbar^2)$ and used $\frac{d}{dh} \VLO = 0$ at $h=\muX$. Note that this is the RG-improved effective potential: the resummation is implicit in the solution for $\muX$.  At NNLO, an infinite number of loops are relevant, even in Landau gauge~\cite{Andreassen:2014eha}.

Using Eq.~\eqref{Vminus} we find that for absolute stability at NLO,
 the Higgs pole mass must satisfy
\be
\frac{\mhpole}{\GeV} > (129.40 \pm 0.58) + 2.26 \, (\frac{\mtpole - 173.34~\GeV}{1.12~\GeV})
\ee
This bound is around 275 MeV lower than the bound from the traditional approach in Landau gauge ($\mhpole > 129.67~\GeV$).
The $\pm 0.58$ is pertubative and $\alpha_s$ uncertainty~\cite{Buttazzo:2013uya}. Since the Higgs mass is known better
than the top mass, it perhaps makes more sense to write the bound as
\be
\frac{\mtpole}{\GeV} < (171.22 \pm 0.28) + 0.12 \, (\frac{\mhpole - 125.14~\GeV}{0.24~\GeV})
\ee

Fig.~\ref{fig:massbound} compares the gauge-dependence of the bound at 1-loop to the LO, NLO and 2-loop bounds. 
For this plot we have taken the U(1) and SU(2) $R_\xi$ gauge parameters equal to $\xi_t$ when $\mu=m_t$ and included their RGE evolution~\cite{Bednyakov:2012rb}. 
All bounds include 2-loop thresholds  and 3-loop running. 
We find that the bound at LO is $\mhpole >129.69~\GeV$ which is 
nearly identical to the Landau gauge 1-loop bound in the traditional approach, 
$\mhpole > 129.70~\GeV$. 
We do not plot the gauge-dependence of the 2-loop bound since we have not computed the gauge-dependent 2-loop potential or the daisy contribution.
That the bound seems to asymptote to a finite value in unitary gauge $(\xi=\infty)$ may be due to much (but not all) of the gauge-dependence being in the $e^{4\Gamma}$ prefactor in Eq.~\eqref{Voldis}
which drops out of the $V=0$ condition.

 \begin{figure}[t]
   \includegraphics[width=0.43\textwidth]{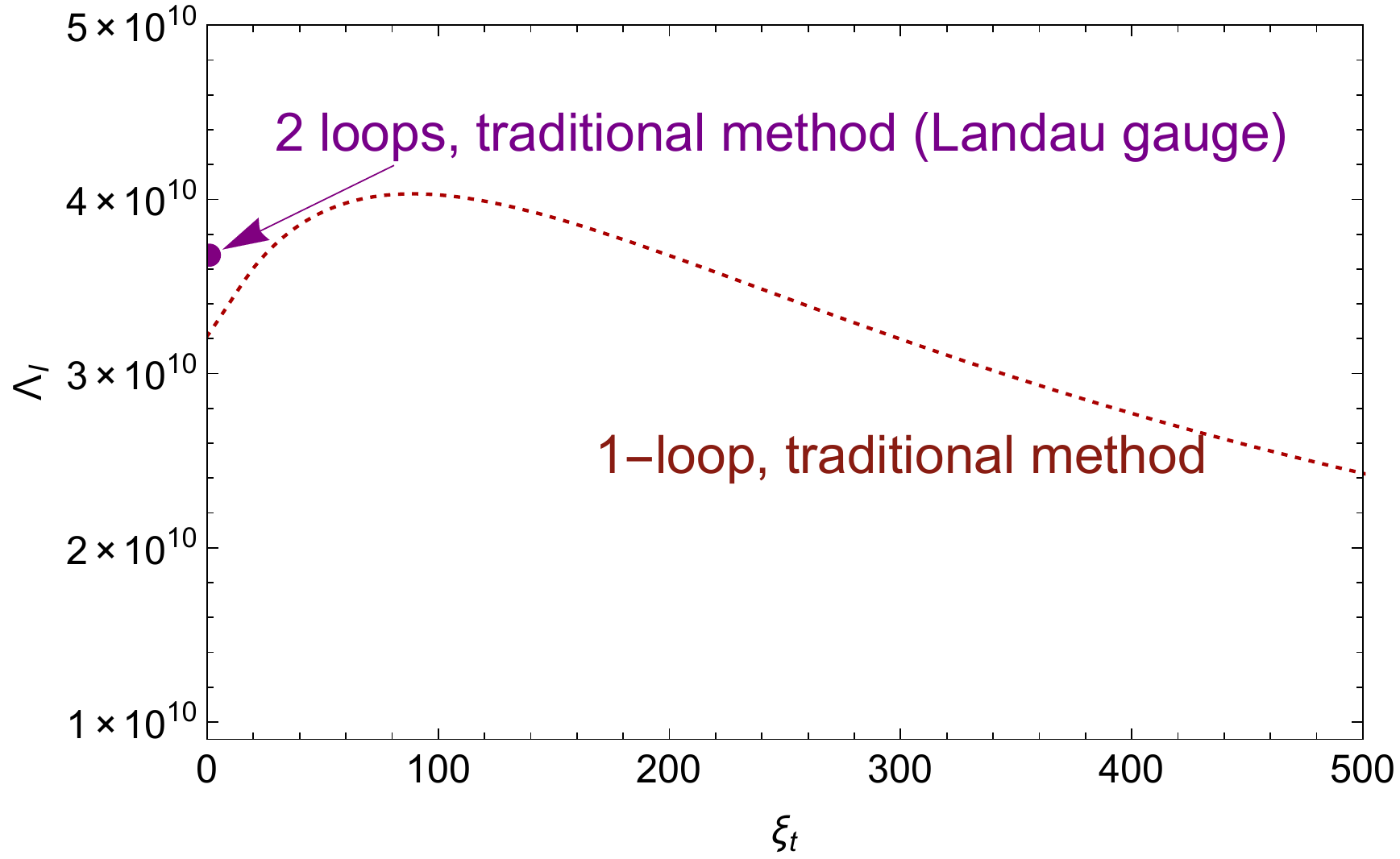}
 \vspace{-2ex}
\caption{Gauge dependence of the instability scale $\Lambda_I$, defined by $V(\Lambda_I)=0$, at 1-loop in the traditional approach. There is no
known way to make this scale gauge-invariant.}
  \label{fig:LambdaI}
 \end{figure}

Fig.~\ref{fig:LambdaI} shows the gauge dependence of the instability scale $\Lambda_I$, defined by $V(\Lambda_I)=0$~\cite{Degrassi:2012ry,Buttazzo:2013uya}, and its Landau-gauge value at 2-loops, including 3-loop resummation in both cases. Since the instability scale is a field value, it is
not obviously physical. We know of no way to compute it in a consistent and gauge-invariant manner. 

\begin{figure}[t]
   \includegraphics[width=0.45\textwidth]{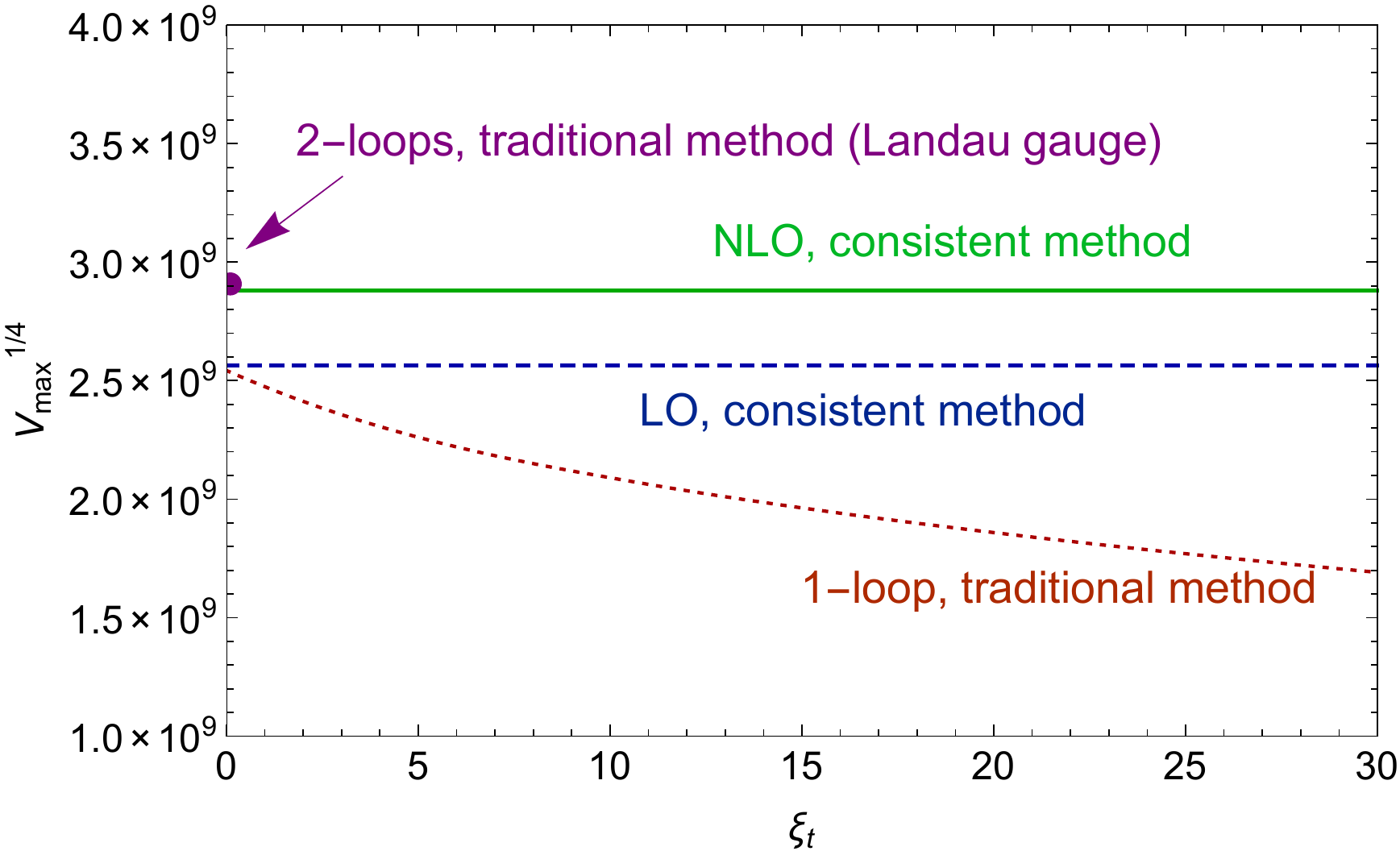}
 \vspace{-2ex}
\caption{Gauge dependence of the SM potential at its maximum with $\mhpole = 125.14~\GeV$ and $\mtpole = 173.34~\GeV$.}
  \label{fig:Vmaxplot}
 \end{figure}

Fig.~\ref{fig:Vmaxplot} shows the value of $\Vmax$ computed by various approaches. We find approximately exponential dependence of $\Vmax$ (and also $\Vmin$) on $\xi_t$ in the traditional approach at 1-loop. Decent fits are
\begin{align}
\nn\\[-12mm]
\\[0mm]
\left(\Vmax^{\text{1-loop, trad.}} \right)^{1/4}&\approx (2.50 \times 10^{9}~\GeV) e^{-0.02 \xi_t+0.0003\xi_t^2}\nn \\
\left(-\Vmin^{\text{1-loop, trad.}} \right)^{1/4}  &\approx   (3.08 \times 10^{29}~\GeV) e^{0.001 \xi_t-0.0001\xi_t^2}\nn
\end{align}
The consistent gauge-invariant values at NLO are
\begin{align}
\left(\Vmax^{\text{NLO}} \right)^{1/4} &= 2.88  \times 10^{9} ~\GeV\label{VminNLO}  \\
\left(-\Vmin^{\text{NLO}} \right)^{1/4} &=  2.40 \times 10^{29} ~ \GeV \nn
\end{align}

Note that $-\Vmin$ corresponds to an energy density well above the Planck scale.
Thus, the potential at the minimum will surely be effected by quantum gravity and possible new physics not included in our calculation.
 Previous analyses have defined stability to be Planck-sensitive if the instability scale $\Lambda_I > \mpl$~\cite{Degrassi:2012ry,Buttazzo:2013uya}.  As we have observed, the instability scale is gauge dependent, so this is not a consistent criterion. An alternative criterion is that new operator, such as 
$\cO_6 \equiv \frac{1}{\LNP^2} h^6$ be comparable to $\Vmin$ when $h= \langle h\rangle$.
Although  $\cO_6$ and $\Vmin$ are gauge-invariant, the value of $\cO_6$ at the field value $h$ where the minimum occurs {\it is} gauge dependent, so this condition is also unsatisfactory.  A consistent and satisfactory criterion was explained in~\cite{Andreassen:2014eha}: the new operator must be added to the classical theory and its effect on $\Vmin$ evaluated. 

Adding $\cO_6$ to the potential, we find that the the potential is still negative at its minimum in the SM  even for operators with very large coefficients. 
For example, taking $\LNP=\mpl = 1.22 \times 10^{19}~\GeV$, we find
that $\muX^{\text{min}} = 6.0 \times 10^{17}~\GeV$ and $\Vmin = - (1.1 \times 10^{17}\, \GeV)^4$.
Comparing to Eq.~\eqref{VminNLO} we see that the energy of the true vacuum is very Planck-sensitive.

More generally, a good fit is given by
\be
\Vmin = - (0.01\, \LNP)^4, \qquad \LNP \gtrsim  10^{12}\,\GeV
\ee
When $\LNP< 3.6 \times 10^{12}~\GeV$, $\Vmin$ becomes positive and for  $\LNP< 3.1 \times 10^{12}~\GeV$ the maximum and minimum disappear. Thus the stability of the Standard Model can be modified by new physics
at the scale $10^{12}~\GeV$. 
 
\begin{figure}[t]
   \includegraphics[width=0.42\textwidth]{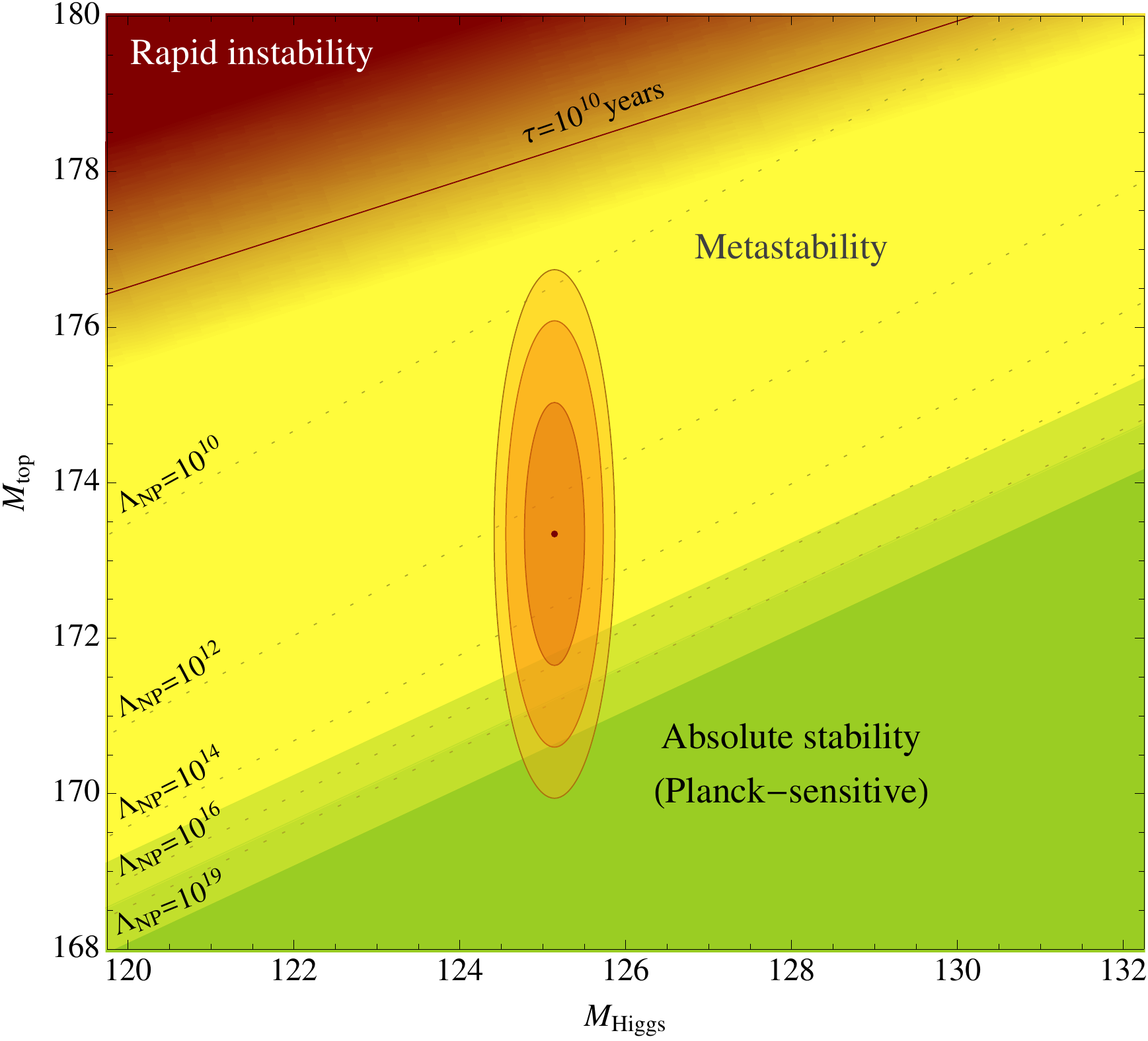}
\caption{
Boundaries of absolute stability (lower band, NLO) and metastability (upper line, LO). 
The thickness of the lower boundary indicates
perturbative and $\alpha_s$ uncertainty. The theoretical uncertainty of the metastability boundary is unknown. 
 The elliptical contours are $68\%$, $95\%$ and $99\%$ confidence bands on the Higgs and top masses: $\mhpole = (125.14 \pm 0.23)~\GeV $ and $\mtpole = (173.34 \pm 1.12)~\GeV$. 
Dotted lines are scales in $\GeV$ at which $\Vmin$ can be lifted positive by new physics.
}
 \vspace{-3ex}
  \label{fig:mhmt}
 \end{figure}

\begin{figure}[t]
   \includegraphics[width=0.42\textwidth]{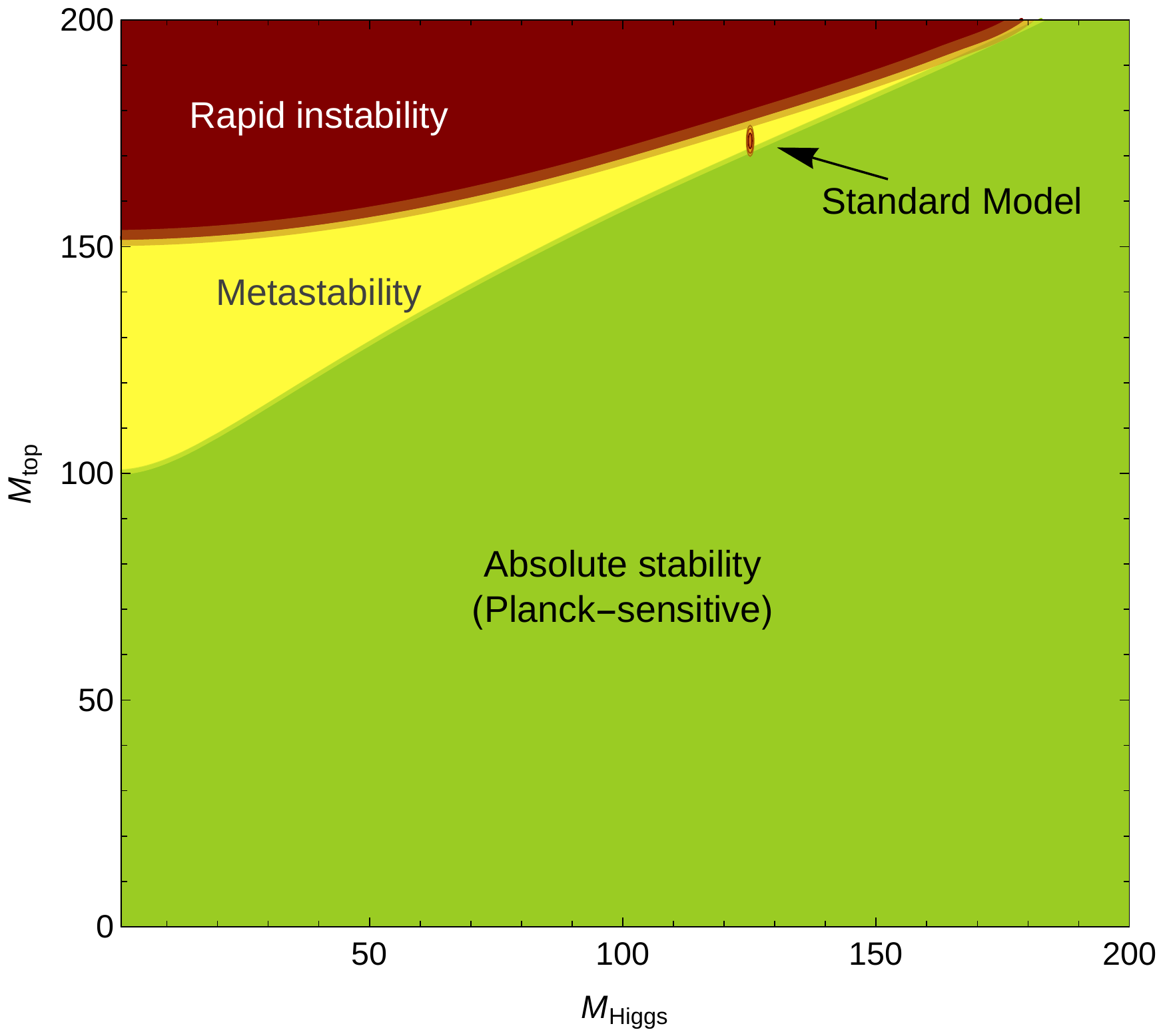}
\caption{
Same as previous figure but zoomed out. Theory uncertainties on the absolute and metastability bounds are not shown.
}
 \vspace{-3ex}
  \label{fig:mhmtbig}
 \end{figure}

If we vary the Higgs and top masses in the Standard Model, we can compute the boundary of absolute stability. 
This bound is shown in Figs.~\ref{fig:mhmt} and~\ref{fig:mhmtbig}. The dotted lines show where $\Vmin$ becomes positive when in the presence of $\cO_6$ for the indicated value of $\LNP$.  
Unexpectedly, we find that three independent conditions (1) that $\Vmin$ goes to zero, (2) that Eq.~\eqref{lambdacond} have no solution,
and (3) that $\Vmin$ goes positive when $\LNP=\mpl$ 
 all give nearly identical boundaries in the $\mhpole$/$\mtpole$ plane. 
Knowing that quantum gravity is relevant at $\mpl$, we should therefore be cautious about giving
too strong of an interpretation of the perturbative absolute stability bound in the SM.
We also show in this plot the metastability bound, that the lifetime of our vacuum be larger than the age of the universe. At lowest order
 this translates
to $\lambda(\frac{1}{R})^{-1} < -14.53 +0.153 \ln [R\,\GeV]$ for all $R$~\cite{Isidori:2001bm}. Since $\lambda(\mu)$ is gauge invariant, so is this criterion. 
Although for the Standard Model this approximation is probably sufficient, it has not been demonstrated that the bound can be systematically improved
in a guage-invariant way~\cite{Strumia:1998nf}.

In this paper, we have only discussed a single physical feature of the effective action: the value of the effective potential at its extrema. There is of course much more content in the effective action, especially when temperature dependence is included. Unfortunately, many uses of the effective action involve evaluating it for particular field configurations, a procedure that has repeatedly been shown to be gauge-dependent. For example, the gauge-dependence of various  quantities associated with the electroweak phase transition were discussed~\cite{Wainwright:2011qy} and various 
predictions of Higgs inflation models~\cite{Barvinsky:2008ia} in~\cite{Cook:2014dga}.

Since observables such as the gravitational wave spectrum or the size of tensor fluctuations in the cosmic microwave background can in principle be predicted within quantum field theory, it should be possible to at least set up such calculations in a way that does not depend on arbitrary gauge or scale choices in the effective action.  Questions which involve other parts of the effective action besides the potential provide new opportunities for cancellation. For example, after the $Z$-factors are added according to the LSZ reduction theorem, $S$-matrix elements calculated from the effective action are appropriately invariant. It would be interesting to see perturbative demonstrations of the gauge-invariance of other derived quantities from the effective action. 

We thank S. Moch, M. Ramsey-Musolf, M. Reece, M. Strassler and A. Strumia for helpful discussions.
The authors are supported in part by grant DE-SC003916 from the Department of Energy. AA is supported in part by the Stolt-Nielsen Fund for Education of the
the American-Scandinavian Foundation and the Norway-America Association. WF is supported in part by the Lord Rutherford Memorial Research Fellowship.

\bibliography{vSM}
\end{document}